\documentclass[pra,aps,twocolumn]{revtex4}
\usepackage{graphicx,epsf,subfigure,amsbsy,amsmath,amsfonts,float}
\usepackage[latin1]{inputenc}

\newcommand{\ket}[1]{\mbox{$|#1\rangle$}}
\newcommand{\bra}[1]{\mbox{$\langle#1|$}}

\begin{document}

\title{Interference Between Competing Pathways in the Interaction of Three-Level Ladder Atoms and Radiation}

\author{Tony Y. Abi-Salloum}

\affiliation{Physics Department, Drexel University,
Philadelphia, Pennsylvania 19104, USA}

\date{\today, {\it Physical Review A}: To be resubmitted.}

\begin{abstract}
In this paper we explore the physical origin of the transparency
induced in absorbing three-level cascade atoms by the simultaneous
interplay of two coherent beams of light. By utilizing the
scattering technique we offer what we believe is a very convincing
physical evidence for the existence, or for the absence, of
quantum interference effects in the Autler-Townes (AT) and
Electromagnetically Induced Transparency (EIT) phenomena.
\end{abstract}

\pacs{PACS numbers: 42.50.Gy}

\maketitle

\section{Introduction}
\label{Introduction}

The interaction of two coherent fields with three-level atoms
features a variety of phenomena which are at the forefront of
research in field of non-linear optics today. A few examples of
some effects include electromagnetically induced transparency
\cite{Fleischhauer:2005}, coherent population trapping
\cite{Arimondo:1996}, line narrowing and lasing without inversion
\cite{Grynberg:1996}. In this article, we highlight first a number
of optical phenomena of interest. Two of the reviewed phenomena
are at the core of this study where quantum interference effects
are investigated.

In 1990 the term ``Electromagnetically Induced Transparency (EIT)"
was introduced by Harris et al. \cite{Harris:1990} whose
theoretical work showed an enhancement in the third-order
susceptibility experienced by a probe field and simultaneously a
decrease in the absorption of the same probe field in a collection
of three-level atoms in a cascade configuration. Many reviews
\cite{Marangos:1997, Harris:1997, Fleischhauer:2005} have been
written about EIT since its emergence. The first experiment
confirming the EIT phenomenon was done in 1991 by Boller,
Imamoglu, and Harris \cite{Boller:1991} in a Strontium lambda
system. Boller and colleagues stated without proving that the
transparency may be interpreted as a combination of the Stark
effect and another interference phenomenon that may in turn be
understood at the level of the dressed states.

In correspondence with the interpretation presented by Boller and
his colleagues, the underlying physical phenomenon responsible for
EIT has been generally assumed to be similar to the physical
origin of the Fano profile \cite{Fano:1961}. In the Fano profile
setting two ionization pathways, one direct and another one
proceeding through an intermediate autoionized state, interfere
leading to a zero transition probability and thus a reduction of
the ionization probability.

Closely related to the EIT phenomenon but one that preceded EIT by
almost 15 years, Coherent Population Trapping (CPT)
\cite{Alzetta-1976, Whitley:1976, Arimondo:1976, Gray:1978} is
another important phenomenon that is featured by three-level
atoms. While EIT is associated with the reduction in the
absorption of the probe field due to the interference between
different excitation pathways, CPT is characterized by the
reduction in spontaneous emission due to the trapping of the
population in the non-coupled state. Other important phenomena,
such as Lasing Without Inversion (LWI) \cite{Grynberg:1996} and
``subluminal" \cite{Harris:1992, Xio:1995, Kasapi:1995, Han:1999,
Kash:1999} and ``superluminal" \cite{Steinberg:1994, Wang:2001}
light, `slow' and `fast' light was also reviewed extensively in
2002 by Boyd \cite{Boyd:2002-Slow-Fast}, are consequences of the
EIT and CPT effects.

A phenomenon which is quite unlike most of what we have surveyed
this far in this paper is the so-called Autler-Townes (dynamic
Stark-Shift) effect \cite{AT:1955}. Even though, the AT effect
resembles the EIT phenomenon in that they both reduce the
absorption of a probe field at or near resonance, AT, studied by
Cohen-Tannoudji \cite{Cohen-Tannoudji:Amazing-Light}, is generally
not associated with interference. However, despite a large volume
of literature devoted to the study of EIT and AT, nowhere has the
manifestation of (or lack therefore) of interference in these two
related systems been explicitly pointed out. This important
difference between the two effects is the subject of investigation
in this paper.

\begin{figure}[htbp]
\centering
\includegraphics[angle=0,width=8.0cm]{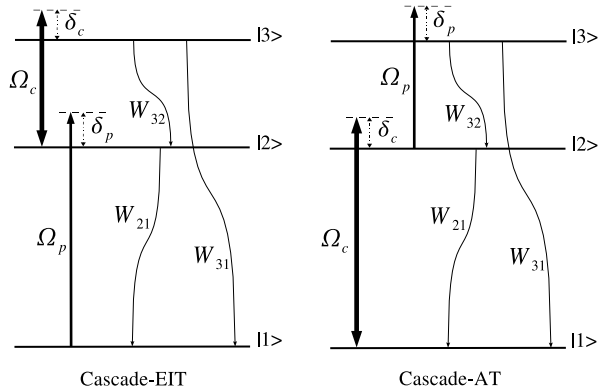}
\caption{Cascade-EIT and Cascade-AT configurations: $\Omega_c$
($\Omega_p$) and $\delta_c$ ($\delta_p$) are the Rabi frequency
and detuning of the coupling (probe) field. $W_{ij}$ is the
spontaneous decay rate from the level $\ket{i}$ to level
$\ket{j}$.}
  \label{Cascade-EIT_AT}
\end{figure}

We begin our investigations by considering two cascade
configurations. We denote the first cascade configuration (shown
on the left in figure \ref{Cascade-EIT_AT}) as Cascade-EIT. This
name, at least at this early point in this paper, recognizes the
experiments \cite{Harris:1991, Li:1995, Banacloche:1995} which
showed EIT in this specific system. Switching the strengths of the
fields leads to another cascade configuration which we denote as
Cascade-AT (shown on the right in Fig. \ref{Cascade-EIT_AT}). At
first, it may seem counter-intuitive that merely switching the
strengths of the involved fields can dramatically affects the
results. However, we stress that experiments show in the
Cascade-EIT case \cite{Jason:2001} a reduction in absorption even
when the coupling field is weak, while experiments show no similar
features in the Cascade-AT case \cite{Feldbaum:2003} in the same
weak field regime. Cascade-EIT and Cascade-AT are the two
configurations which we have selected for our in-depth analysis of
the similarities and, specially, differences between the EIT and
the AT effects.

Following Lounis and Cohen-Tannoudji \cite{Lounis:1992} we
interpret the absorption of a probe photon as a scattering process
induced by the atom while interacting with several pump (coupling)
photons. We calculate the corresponding scattering amplitude and,
if they exist, we identify the multiple physical paths followed by
the system as it evolves to its final state. The existence of
compatible multiple paths in the transition amplitude results in
quantum interference effects exhibited by the transition
probability. In 1993, Grynberg and Cohen-Tannoudji
\cite{Cohen-Tannoudji:1993}, using the same scattering technique
\cite{Lounis:1992}, traced the physical origin of gain in the
central resonance of the Mollow absorption spectrum. The
scattering technique in the dressed states picture was also used
to study the Vee system \cite{Grynberg:1996}.

In the following section of this paper, two coupling field regimes
are highlighted. Unlike the case in the strong coupling field
regime where the two cascade systems display similar probe
absorption spectra, in the weak field regime, which will be
extensively studied in this paper, the two configurations present
a critical difference. In section \ref{Technique-Scattering}, we
survey the mathematical tools which are needed to calculate the
probability that a system evolves from a given initial to a final
state. For certain crucial technical reasons discussed in the
introduction of section \ref{Bare-States-Picture-Scattering}, only
the Cascade-EIT configuration can be studied in the bare states
picture. It is in this same section where we show the existence of
two excitation-emission pathways in the process of absorption of a
probe photon in the Cascade-EIT case. The results are clarified in
the low saturation limit where we prove that the two pathways
compete and interfere. Section
\ref{Dressed-States-Picture-Scattering} studies the scattering of
a probe photon in the dressed states picture only in the low
saturation limit due to physical requirements of the technique
which will be discussed in the introduction of the section. Both
configurations are studied. The expected absence of interfering
pathways in the Cascade-AT case is verified. Also, the results
found in the Cascade-EIT case are consistent with the ones found
in the bare state picture.

\section{Macroscopic Differences and Similarities}
\label{MDaS}

\begin{figure}[htbp]
  \centering
\includegraphics[angle=0,width=8cm]{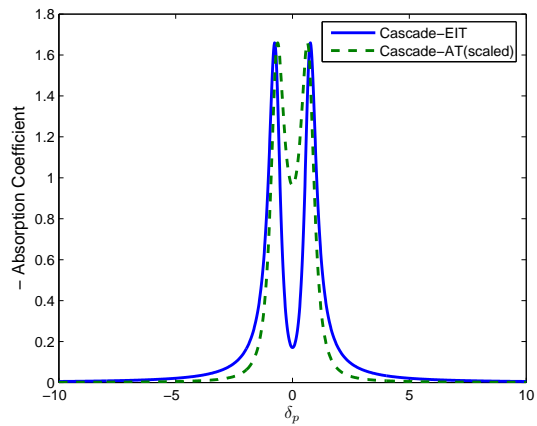}
  \caption{Negative absorption coefficient of the probe field in the Cascade-EIT case (solid line)
  and Cascade-AT case (dashed line) function of the probe detuning.
  Theoretical values: $\gamma_{12}$=0.5, $\gamma_{13}$=0.105, $\gamma_{23}$=0.605, $\delta_c$=0, $\Omega_c$=1.5.}
  \label{EIT_AT_abs}
\end{figure}

The probe absorption spectrum (solid line in figure
\ref{EIT_AT_abs}) of the Cascade-EIT configuration, which is
proportional to the imaginary part of the density matrix element
$\rho_{21}$ \cite{Banacloche:1995},

\begin{equation}
\rho_{21} \;\propto\; -i \;
\dfrac{\gamma_{13}-i(\delta_p+\delta_c)}{
    {|\Omega_c|^2 \over 4} + [\gamma_{12}-i \delta_p][\gamma_{13}-i(\delta_p+\delta_c)]},
\label{abs_EIT_copied}
\end{equation}

and the probe absorption spectrum (dashed line in figure
\ref{EIT_AT_abs}) of the Cascade-AT configuration, which is
proportional to the imaginary part of $\rho_{32}$
\cite{Vemuri:1995},

\begin{eqnarray}
\rho_{32} &\propto& i{ {|\Omega_c|^2\over 4} \over \gamma_{12}^2
 + \delta_c^2+2{|\Omega_c|^2\over 4}}\;\times\\
& &{\gamma_{23}+i \delta_p \over
[\gamma_{13}+i(\delta_p+\delta_c)][\gamma_{23}+i
\delta_p]+{|\Omega_c|^2 \over 4}},\nonumber
\end{eqnarray}

display similar features, mainly a reduction in absorption in the
expected maximum absorptive probe field detuning range in the
absence of the coupling field. Note that we changed several
notations such as detunings, decay rates, density matrix elements,
and Rabi frequencies, redefined the coupling Rabi frequency,
included the spontaneous decay rate $W_{31}$, and also dropped
multiplication factors for the sake of consistency between the two
references \cite{Banacloche:1995, Vemuri:1995} and our work. All
the previously used variables will be defined in the following
sections when needed.

\begin{figure}[htbp]
  \centering
\includegraphics[angle=0,width=8cm]{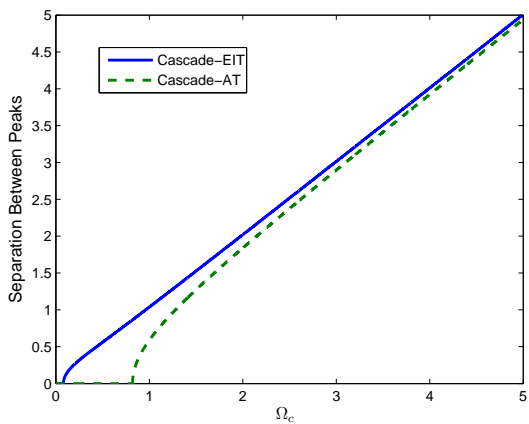}
  \caption{Separation between the peaks of the probe absorption line in the Cascade-EIT case (solid line)
  and Cascade-AT case (dashed line) function of the coupling field Rabi frequency.
  Theoretical values: $\gamma_{12}$=0.5, $\gamma_{13}$=0.105, $\gamma_{23}$=0.605,
  $\delta_c$=0.}
  \label{Separation_EIT_AT}
\end{figure}

Figure \ref{Separation_EIT_AT} is a plot of the separation between
the peaks of the absorption spectra at resonance ($\delta_c$ = 0)
function of the coupling field Rabi frequency. Monitoring the
separation between the peaks reveals two regimes. The strong field
regime ($\Omega_c \gg \gamma_{ij}$) has been studied by
Cohen-Tannoudji \cite{Cohen-Tannoudji:Book,
Cohen-Tannoudji:1977-3,Cohen-Tannoudji:Amazing-Light} and
co-authors. The authors study the absorption and fluorescence
spectra of a variety of three-level systems using the dressed-atom
approach \cite{Cohen-Tannoudji:RF} in the secular limit (the
effective Rabi frequency of the system, which is related to the
strengths of the different used fields, is much greater than the
atomic decay rates) which in our case is equivalent to the strong
coupling field regime. One of the common messages across the
different studies presented by Cohen-Tannoudji and co-authors is
that in the three-level system case and in the secular limit the
absorption line of the probe field splits into two Lorentzian
lines which at resonance are separated by the coupling field Rabi
frequency. This split in the absorption line is a consequence of
the AT effect \cite{Cohen-Tannoudji:Amazing-Light} which at least
in the secular limit is not associated with interference. Figure
\ref{Separation_EIT_AT} displays the AT linear separation in the
strong coupling field range of Rabi frequency.

The linear AT separation displayed by both cascade configurations
shows that the AT effect dominates all other possibly existing
phenomena in the strong coupling field regime. As the coupling
field strength decreases it is obvious that in the Cascade-AT case
the dip fills up quickly as the two Lorentzians start overlapping
while in the Cascade-EIT case the transparency persists and the
separation even goes above the linear line and does not vanish
until the coupling field is almost turned off. This macroscopic
evident difference between the two cascade configurations, which
we believe is a manifestation of the microscopic difference
between the two phenomena AT and EIT, is the point of
investigation in this paper. For this very clear reason we focus
in this work on the weak coupling field regime which could be
achieved by either one of the two conditions $\Omega_c \ll
\gamma_{ij}$ or $\Omega_c \ll \delta_c$ which are known as the low
saturation limit. We note that the low saturation condition
$\Omega_c \ll \gamma_{ij}$ varies between the different
configurations and will be given later in the paper.

\section{Technique}
\label{Technique-Scattering}

In this section we provide an outline of the general elements of
the scattering technique \cite{Cohen-Tannoudji:Book} that is
adopted in this paper. The transition and probability amplitudes,
which are related to the probability of absorption are defined and
calculated non perturbatively in the first subsection
\ref{Transition and Probability Amplitudes} and then rederived in
the second subsection \ref{Resolvent Operator} after the
introduction of an operator called the Resolvent.

\subsection{Transition and Probability Amplitudes}
\label{Transition and Probability Amplitudes}

A system in an initial state $\ket{i}$ at a time $t_i$ has a
probability $\mathcal{P}_{fi}(t_f,t_i)$ of performing a transition
to a final state $\ket{f}$ at a later time $t_f$. The system
evolves between the two times, $t_i$ and $t_f$, according to the
unitary transformation
\begin{equation}
\ket{f} \;=\; U(t_f,t_i) \ket{i},
\label{evolution-Scattering}\end{equation}
where the evolution operator, $U(t_f,t_i)$, is defined as
\begin{equation}
U(t_f,t_i)\;=\; e^{-i H (t_f-t_i)/ \hbar},
\end{equation}
and depends on the total Hamiltonian H.

The probability amplitude that the system will go from the initial
state $\ket{i}$ to the final state $\ket{f}$ is defined as
\begin{equation}
\mathcal{S}_{fi} \;\equiv\; U_{fi} \;=\; \bra{f}U(t_f,t_i)\ket{i}.
\label{S_fi-definition-technique-Scattering}
\end{equation}

The transition probability, $\mathcal{P}$, is equal to the modulus
squared of the the probability amplitude, $\mathcal{S}_{fi}$,
\begin{equation}
\mathcal{P}_{fi}(t_f,t_i) \;=\; |\mathcal{S}_{fi}|^2.
\label{P-original-definition-Scattering}
\end{equation}

After substituting equation \ref{evolution-Scattering} into the Schr\"odinger
equation for the final state we solve for the evolution operator, $U(t_f,t_i)$,
which after a transformation into the interaction
picture and a substitution into equation \ref{S_fi-definition-technique-Scattering} leads to
\begin{equation}
\mathcal{S}_{fi} \;\;\simeq\;\; \delta_{fi} - 2 \pi i
\delta^{(T)}(E_f-E_i) \lim_{\eta \rightarrow 0_{+}}
\mathcal{T}_{fi}, \label{S_fi-last-11-Scattering}\end{equation}
where we defined the transition amplitude, $\mathcal{T}_{fi}$ (not
to be confused with the probability amplitude $\mathcal{S}_{fi}$),
as
\begin{equation}
\mathcal{T}_{fi} \;=\; \bra{f}V\ket{i} + \bra{f}V
\dfrac{1}{E_i+i\eta-H} V\ket{i},
\label{T_fi-explicit-V-Scattering}\end{equation}
V being the interaction part of the Hamiltonian, $\eta$ being a positive real number introduced
by the identity
\begin{eqnarray}
e^{i E (t_2-t_1)/ \hbar} \Theta(t_2-t_1) &=& \nonumber \\
& & \hspace{-2cm} \lim_{\eta \rightarrow 0_{+}} -\dfrac{1}{2\pi i}
\int^{+\infty}_{-\infty} \dfrac{e^{-i E (t_2-t_1)/
\hbar}}{E+i\eta-E_k} dE,
\end{eqnarray}
where $\Theta(t_2-t_1)$ is the Heaviside function. We also used in
equation \ref{S_fi-last-11-Scattering} the diffraction function
$\delta^{(T)}(E_f-E_i)$, which is defined as
\begin{equation}
\delta^{(T)}(E_f-E_i) \;=\; \dfrac{1}{\pi}
\dfrac{\sin\left[(E_f-E_i)T/2\hbar\right]}{E_f-E_i},
\end{equation}
where T is the duration of the interaction.

\subsection{Resolvent Operator}
\label{Resolvent Operator}

We introduce the resolvent operator of the total Hamiltonian, H, in the form
\begin{equation}
G(z) \;=\; \dfrac{1}{z-H},
\label{G-definition-Secular}\end{equation}
and reduce equation \ref{T_fi-explicit-V-Scattering} to
\begin{equation}
\mathcal{T}_{fi} \;=\; \bra{f}V\ket{i}+\bra{f}V G(E_i+i\eta)
V\ket{i}. \label{T-fi-general-Secular}\end{equation}

For use later on, we can re-write equation
\ref{T-fi-general-Secular} by introducing the identity operator
$\hat{1} \equiv \sum_k \ket{k}\bra{k} $, so that
\begin{eqnarray}
\mathcal{T}_{fi} &=& \bra{f}V\ket{i} + \bra{f}V \hat{1}\dfrac{1}{E_i+i\eta-H} \hat{1}V\ket{i},\\
&=& \bra{f}V\ket{i} + \sum_k \sum_l V_{fk} V_{li} \bra{k}
G(E_i+i\eta) \ket{l}.
\end{eqnarray}

In the cases of interest where $\ket{f} \neq \ket{i}$ the
probability $\mathcal{P}_{fi}$ (Eq.
\ref{P-original-definition-Scattering}) is given by
\begin{equation}
\mathcal{P}_{fi} \;=\; 4 \pi^2
\left[\delta^{(T)}(E_f-E_i)\right]^2 \; |\lim_{\eta \rightarrow
0_{+}}\mathcal{T}_{fi}|^2,
\end{equation}
which implies that
\begin{eqnarray}
\hspace*{-0.3cm} \mathcal{P}_{fi} &=& \pi^2  \left[\delta^{(T)}(E_f-E_i)\right]^2 \times \nonumber\\
& & \hspace*{-0.5cm} |V_{fi}+\sum_{k,l} V_{fk} V_{li} \bra{k}
\lim_{\eta \rightarrow 0_{+}}G(E_i+i\eta) \ket{l}|^2.
\end{eqnarray}

The states $\ket{k}$ and $\ket{l}$ are discrete intermediate
states of the scattering process between the initial and final
states, $\ket{i}$ and $\ket{f}$. The system is not observed at the
intermediate states, but the the probability amplitude between the
initial and final states is the sum over all probability
amplitudes of all the intermediate states.

The matrix element $\bra{k} G(E_i+i\eta) \ket{l}$ of the resolvent
operator, G, is the matrix element of G projected onto the two
subspaces to which the states $\ket{k}$ and $\ket{l}$ belong. For
example, if the two intermediate states $\ket{k}$ and $\ket{l}$
belong to the same subspace $\mathcal{E}_o$ which is made of the
set of quasi-degenerate and coupled states $\left\lbrace
\ket{\varphi_{1}},...\;,\ket{\varphi_{n}}\right\rbrace$, we have
\begin{equation}
G_{kl}(E_i+i\eta)=\bra{k} PG(E_i+i\eta)P \ket{l},
\end{equation}
where the projector P is defined as
\begin{equation}
P \;=\; \sum_i^n \ket{\varphi_i}\bra{\varphi_i}.
\end{equation}

The other possible case is when only one of the intermediate
states, $\ket{l}$ for example, belongs to the subspace
$\mathcal{E}_o$ while the second state, $\ket{k}$, belongs to the
supplementary subspace of $\mathcal{E}_o$, $\mathcal{L}_o$. In
this case we have

\begin{equation}
G_{kl}(E_i+i\eta)=\bra{k} QG(E_i+i\eta)P \ket{l},
\label{G-kl-QGQ-Secular}\end{equation}
where Q is the supplementary projector of P defined as
\begin{equation}
Q \;=\; \hat{1} - P.
\end{equation}

After some calculations we find that
\begin{eqnarray}
P G(z)P &=& \dfrac{P}{z-P H_o P -P R(z) P},
\label{PG(z)P-R-Secular}\\
QG(z)P &=& \dfrac{Q}{z-QHQ}\;V\; P G(z)P.
\label{QG(z)P-R-Secular}
\end{eqnarray}
where $H_o$ is the unperturbed Hamiltonian, $H$ is the total
Hamiltonian ($H=H_o+V$), and R(z) is the level shift operator
defined as
\begin{equation}
R(z) \;=\; V + V \dfrac{Q}{z-QHQ} V.
\label{R-definition-Secular}\end{equation}

\section{Bare States Picture}
\label{Bare-States-Picture-Scattering}

The probability that a system will evolve from an initial to a
final state is independent of the picture in which the
calculations are carried out. The different pictures only effect
the intermediate states associated with the pathways followed by
the system as it evolves from the initial to the final state. Some
pictures, the dressed states picture for example, pose specific
restrictions to the application of the scattering technique. This
fact will be addressed further in section
\ref{Dressed-States-Picture-Scattering} where the dressed states
picture will be used.

In this section we adopt the bare states picture seeking an
understanding of the possible pathways followed by the atom
through intermediate bare states in the process of scattering one
probe photon. We consider the physically realistic situation where
the system, in this case the bare atom, is initially in a
quasi-stable state. At the end of the scattering process the
system must also exist in a quasi-stable state. If this was not
the case, the evolution would not yet be over. In this section as
in the subsequent sections, we will always only consider initial
and final states that are quasi-stable.

In a cascade system the excited atomic states $\ket{2}$ and
$\ket{3}$ are not stable because of their spontaneous decay out of
these states. This leaves the ground state, $\ket{1}$, as the only
bare atomic state which can be used as an initial and final state
for the scattering process. In the Cascade-EIT configuration,
unlike the Cascade-AT configuration, the state $\ket{1}$ is
coupled by a weak field (probe) to state $\ket{2}$. This fact
keeps the state $\ket{1}$ stable. This situation is quite
different from the Cascade-AT case where state $\ket{1}$ is
coupled to state $\ket{2}$ by a strong field (coupling). The fact
that the strong coupling field forces the atom to oscillate
between the two states $\ket{1}$ and $\ket{2}$ with a large Rabi
frequency makes the state $\ket{1}$ unstable. The problem of not
having a bare stable state in the Cascade-AT configuration
eliminates the possibility of applying the scattering technique to
this configuration in the bare states picture. We study in this
section only the Cascade-EIT configuration. In the next section,
instead, we show that one of the dressed states in either
configuration is quasi-stable and this allows the study of the
Cascade-AT configuration in the dressed state picture.

The initial setting of the Cascade-EIT system of interest involves
the atom in the atomic ground state $\ket{1}$ interacting
simultaneously with the coupling field, having $N_c$ photons in
its mode, one probe photon, and zero photons in all the different
vacuum modes, which we label by a subindex j. Thus, the initial
state has the form
\begin{equation}
\ket{i} \;=\;
\ket{1;(1)_p,(N_c)_c,(0)_j}.\label{initial-state-Bare-EIT-Scattering}
\end{equation}

Considering only one photon in the probe field corresponds to the
simplest scattering process, where one probe photon is absorbed
followed at a later time by the emission of a new photon in one of
the vacuum modes. We did examine the case where $N_P$ ($N_P$ $>$
1) photons exist in the probe field. We found that the inclusion
of more than one probe photon is associated with higher order
interactions which is out of the scope of this paper because of
their low orders of magnitudes.

Based on the scattering process and the requirement that the state
be quasi stable, the final state corresponds to the situation
where the atom is in its ground state with $N_c$ photons in the
coupling field, no photons in the probe field, and one photon of
frequency $\omega$ in the corresponding vacuum mode (with all
other vacuum modes having zero photons). This state is given by
\begin{equation}
\ket{f} \;=\;
\ket{1;(0)_p,(N_c)_c,(1)_{\omega}}.\label{final-state-Bare-EIT-Scattering}
\end{equation}

After using the appropriate initial (Eq.
\ref{initial-state-Bare-EIT-Scattering}) and final (Eq.
\ref{final-state-Bare-EIT-Scattering}) sates, the transition
amplitude, $\mathcal{T}_{fi}$ (Eq. \ref{T-fi-general-Secular}),
reduces to
\begin{equation}
\mathcal{T}_{fi} \;=\; \dfrac{\hbar^2 \Omega\Omega_p}{4} \bra{
\varphi_2} G(E_i+i\eta)\ket{\varphi_2},
\label{T_fi-Bare-EIT-Scattering}\end{equation}
where $\Omega_p$ and $\Omega$ are respectively the Rabi
frequencies of the probe and the non-zero vacuum field defined in
this case as
\begin{subequations}
\begin{eqnarray}
\Omega_p &=& -2\mu_{12}\sqrt{\frac{\hbar\omega_p}{2\epsilon_o
L^3}},\\
\Omega &=&-2\mu_{12}\sqrt{\frac{\hbar\omega}{2\epsilon_o L^3}},
\end{eqnarray}
\label{rabi_frequency_definition}
\end{subequations}
where $L^3$ is equal to the volume of the cavity, and where the
discrete state $\ket{\varphi_2}$ is defined as
\begin{equation}
\ket{\varphi_2} \;=\;
\ket{2;(0)_p,(N_c)_c,(0)_j}.\label{varphi-2-Scattering}
\end{equation}

By setting the energy of the state $\ket{\varphi_2}$ as the energy
reference, $E_{\varphi_2}=0$, and after defining the state
$\ket{\varphi_3}$ as
\begin{equation}
\ket{\varphi_3} \;=\;
\ket{3;(0)_p,(N_c-1)_c,(0)_j},\label{varphi-3-Scattering}
\end{equation}
we find that the states $\ket{i}$ and $\ket{\varphi_3}$ have the
energies
\begin{subequations}
\begin{eqnarray}
E_i &=& \hbar \delta_p,\\
E_{\varphi_3} &=& -\hbar \delta_c,
\end{eqnarray}
\label{Energies-EIT-Bare-Secular}\end{subequations}
where $\delta_p=\omega_p-\omega_{12}$ and
$\delta_c=\omega_c-\omega_{23}$, and are quasi-degenerate with the
state $\ket{\varphi_2}$ (the quasi-degenerate results from the
smallness of the detuning parameters relative to all the other
frequencies of the problem).

In the infinite volume limit, $L\rightarrow \infty$, we find that
the coupling between the states $\ket{\varphi_3}$ and
$\ket{\varphi_2}$,
\begin{equation}
\bra{\varphi_2}V\ket{\varphi_3} \;=\; \hbar \dfrac{\Omega_c}{2}\;
\stackrel{L\rightarrow \infty}{\longrightarrow \hspace*{-0.6cm}
\times}\;0,
\end{equation}
where $\Omega_c$ is the Rabi frequency of the coupling field
defined as $\Omega_c \;=\; -2\mu_{23}\sqrt{\frac{\hbar\omega_c N_c
}{2\epsilon_o L^3}}$, remains finite.

Unlike the coupling between the states $\ket{\varphi_3}$ and
$\ket{\varphi_2}$, the coupling between the two states
$\ket{\varphi_2}$ and $\ket{i}$,
\begin{equation}
\bra{\varphi_2}V\ket{i} \;=\; \hbar \dfrac{\Omega_p}{2} \;
\stackrel{L\rightarrow \infty}{\longrightarrow}\; 0,
\end{equation}
vanishes in the infinite volume limit. This fact eliminates the
initial state $\ket{i}$ from the subspace of the state
$\ket{\varphi_2}$.

In this case the matrix element $\bra{\varphi_2}
G(E_i+i\eta)\ket{\varphi_2}$ is the element of the operator G
projected onto the subspace $\mathcal{E}_o$ defined as
\begin{equation}
\mathcal{E}_o=\left\lbrace \ket{\varphi_2}, \ket{\varphi_3}
\right\rbrace. \label{E_o-23-Scattering}
\end{equation}

Using equation \ref{PG(z)P-R-Secular}, the projection of the
resolvent operator, $G(E_i+i\eta)$, into the subspace
$\mathcal{E}_o$, leads to
\begin{eqnarray}
& & \hspace*{-0.7cm} \lim_{\eta \rightarrow 0_{+}}PG(E_i+i\eta)P\;= \nonumber\\
 & & \hspace*{-0.3cm} \dfrac{\hbar}{D}\left(\begin{array}{cc}
\delta_p+i W_{21}/2 & -\Omega_c/2\\
-\Omega_c/2 & \delta_p+\delta_c+i(W_{32}+W_{31})/2
\end{array}\right),
\label{PG+P-Bare-EIT-Scattering}\end{eqnarray}
where D is the determinant of the matrix
$(PGP)^{-1}$ and is given by
\begin{equation}
D \;=\; \hbar^2 \left(\delta_p+\delta_c+i\gamma_{13}\right)
\left(\delta_p+i\gamma_{12}\right)-\hbar^2\dfrac{\Omega^2_c}{4},
\label{D-one-piece-Scattering}
\end{equation}
where $W_{ij}$ is the spontaneous decay rate from level i to level
j (i,j={2,3}) and $\gamma_{ij}$ is the corresponding polarization
decay rate, which in free space is given by
\begin{equation}
\gamma_{ij}=\sum_{k=1}^3\left(W_{ik}+W_{jk}\right).
\label{polarization_general_eq}
\end{equation}

\subsection{Resonances}

The determinant D (Eq. \ref{D-one-piece-Scattering}) can be written explicitly in terms of the eigenvalues
of the matrix $(PGP)^{-1}$, i.e.
\begin{equation}
D = \hbar^2(\delta_p - Z_{II}) (\delta_p - Z_{III}),
\end{equation}
where the eigenvalue, $Z_{II}$ and $Z_{III}$ are given by
\begin{subequations}
\begin{eqnarray}
2 Z_{II} &=& - (\delta_c + i\gamma_{23}) + \sqrt{(\delta_c+i\gamma_{13}-i\gamma_{12})^2+\Omega^2_c},\nonumber\\
& & \\
2 Z_{III} &=& - (\delta_c + i\gamma_{23}) - \sqrt{(\delta_c+i\gamma_{13}-i\gamma_{12})^2+\Omega^2_c}.
\label{Z_III-original-Scattering}\nonumber\\
\end{eqnarray}
\label{Z_II-Z_III-original-Scattering}\end{subequations}

After writing the transition amplitude (Eq.
\ref{T_fi-Bare-EIT-Scattering}) in terms of the eigenvalues we
obtain
\begin{eqnarray}
\mathcal{T}_{fi} &=& \dfrac{\hbar \Omega \Omega_p}{4(Z_{II} - Z_{III})} \times \nonumber\\
& &\left( \dfrac{Z_{II}+\delta_c+i\gamma_{13}}{\delta_p -Z_{II}} -
\dfrac{Z_{III}+\delta_c+i\gamma_{13}}{\delta_p -Z_{III}}\right).
\label{T_fi-Z_II-Z_III-Scattering}
\end{eqnarray}

Because the transition amplitude is the sum of two complex
numbers, the scattering process may be characterized by an
interference between two possible evolution pathways followed by
the atom during the scattering process. Each of the two complex
numbers is associated with an intermediate state, or a sequence of
intermediate states. We denote these pathways as first and second
resonance and, thus, write equation
\ref{T_fi-Z_II-Z_III-Scattering} in the form
\begin{equation}
\mathcal{T}_{fi} \;=\; \rm{1^{st} \; Resonance} + \rm{2^{nd} \;
Resonance}.
\end{equation}

We note here that what is as important as the existence of two
resonances is the relative order of magnitude between the
amplitudes of these resonances. This business of relative orders
will be clarified in the next subsection in the low saturation
limit where the interference effect is significant.

When the coupling field is turned off, $\Omega_c = 0$, the
eigenvalues, $Z_{II}$ and $Z_{III}$, approach their unperturbed
values
\begin{subequations}
\begin{eqnarray}
Z_{II} &\rightarrow& - i \gamma_{12},\label{Z_II-aprox}\\
Z_{III} &\rightarrow& - \delta_c - i\gamma_{13}.
\end{eqnarray}
\end{subequations}

Consistently, we expect that the eigenstates
$\bar{\ket{\varphi_{2}}}$ and $\bar{\ket{\varphi_{3}}}$ which
correspond to the eigenvalues $Z_{II}$ and $Z_{III}$ approach the
unperturbed states $\ket{\varphi_{2}}$ and $\ket{\varphi_{3}}$ in
the absence of the coupling field ($\Omega_c = 0$). Note that the
bars in $\bar{\ket{\varphi_{2}}}$ and $\bar{\ket{\varphi_{3}}}$
are added to distinguish the eigenstates from their corresponding
unperturbed states $\ket{\varphi_{2}}$ and $\ket{\varphi_{3}}$.

Based on the understanding of equation \ref{Z_II-aprox} we set the eigenvalue $Z_{II}$ in the form
\begin{equation}
Z_{II} = - i\gamma_{12} -\delta'_c+i\gamma'_c,
\label{Z_II-approximate-Scattering}\end{equation}
where $\delta'_c$ and $\gamma'$, which will be derived later on in
this paper, are correction terms which are consequences of the
presence of the coupling field.
In addition, after using the conservation of the trace we obtain
\begin{equation}
Z_{III} = -\delta_c - i\gamma_{13} +\delta'_c-i\gamma'_c.
\label{Z_III-approximate-Scattering}\end{equation}

The corrections to the energy and radiative broadening of levels
$\ket{2}$ and $\ket{3}$ due to the existence of the coupling field
can be understood in the following way. The term $-\delta'_c$
($\delta'_c$) represents the light shift of level $\ket{2}$
($\ket{3}$). Similarly, the term -$\gamma'_c$ ($\gamma'_c$) is the
radiative correction of the unperturbed level $\ket{2}$
($\ket{3}$).

The first resonance is centered at
$\delta_p=\Re(Z_{II})=-\delta'_c$, which implies that $\hbar
\omega_p = \hbar \omega_{21}-\hbar \delta'_c$, which is the
optical resonance between level $\ket{1}$ and the shifted level
$\ket{2}$. This optical resonance has a width of
$\gamma_{12}-\gamma'_c$ which approaches $\gamma_{12}$ when
$\Omega_c$ tends to zero.

The second resonance is centered at
$\delta_p=\Re(Z_{III})=-\delta_c+\delta'_c$ which is equivalent to
$\hbar \omega_p+\hbar \omega_c = \hbar \omega_{31}+\hbar
\delta'_c$. This resonance corresponds to the Raman resonance
condition between the light-shifted level $\ket{3}$ and level
$\ket{1}$ and has a linewidth of $\gamma_{13}+\gamma'_c$.

\subsection{Low Saturation Limit}
\label{Bare-States-Picture-lsl-Scattering}

In this section we consider the low saturation limit in which the
interference effect is not dominated by any other phenomena such
as the AT one. In this case the level shifts and the linewidths
corrections acquire more transparent forms and the pathways
corresponding to the different resonances become obvious.

In the low saturation limit, $\Omega^2_c /
(\gamma_{12}-\gamma_{13})^2 \ll 1$ or $\Omega^2_c / \delta^2_c \ll
1$, equation \ref{Z_III-original-Scattering} reduces to
\begin{equation}
Z_{III} \;=\; -\delta_c -i\gamma_{13}\;- \dfrac{\Omega^2_c/4}{\delta_c+i(\gamma_{13}-\gamma_{12})},
\end{equation}
which after comparison with equation \ref{Z_III-approximate-Scattering} leads to
\begin{subequations}
\begin{eqnarray}
\delta'_c &=& -\delta_c \; \dfrac{\Omega^2_c/4}{\delta^2_c+(\gamma_{13}-\gamma_{12})^2},\\
\gamma'_c &=& -(\gamma_{13}-\gamma_{12}) \;
\dfrac{\Omega^2_c/4}{\delta^2_c+(\gamma_{13}-\gamma_{12})^2}.
\end{eqnarray}
\label{delta-gamma-correction-approx-Scattering}\end{subequations}

In the low saturation limit the transition amplitude (Eq.
\ref{T_fi-Z_II-Z_III-Scattering}) takes the form
\begin{equation}
\mathcal{T}_{fi} = \dfrac{\hbar \Omega \Omega_p}{4}  \left(
\dfrac{1}{\delta_p-Z_{II}} + \left[
\dfrac{-\Omega_c/2}{\delta_c+i(\gamma_{13}-\gamma_{12})}\right]^2
\dfrac{1}{\delta_p-Z_{III}} \right),
\label{T_fi-Bare-EIT-before-aprox-Scattering}\end{equation}
where the two terms between the parentheses correspond to the
first and second resonances.

The first resonance is given explicitly by
\begin{equation}
\rm{1^{st} \; Resonance} \;=\;  \dfrac{\hbar \Omega \Omega_p}{4}
\dfrac{1}{\delta_p+\delta'_c+i(\gamma_{12}-\gamma_c')},
\end{equation}
where $\gamma'_c << \gamma_{12}$. We also consider the limit
$\delta'_c << \delta_p$ which yields the final result
\begin{equation}
\rm{1^{st} \; Resonance} \;\approx\; \dfrac{\hbar \Omega }{2} \;
\dfrac{1}{\hbar(\delta_p+i\gamma_{12})}\; \dfrac{\hbar \Omega_p
}{2}. \label{first_resonance_EIT_bare_final_version}
\end{equation}

\begin{figure}[htbp]
  \centering
\includegraphics[angle=0,width=6.0cm]{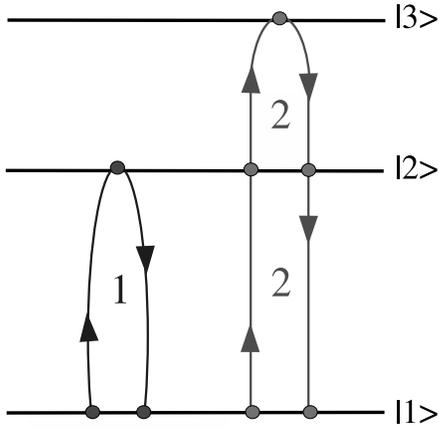}
  \caption{Resonances of the Cascade-EIT configuration in the bare states picture.}
  \label{Resonances-Bare-EIT-Scattering}
\end{figure}

The first resonance is the product of three factors. Starting from
the right hand side, the factor $\hbar \Omega_p/2$ describes the
absorption process of the probe photon. This absorption process
leaves the atom in the state 2 which has an energy $\hbar
\delta_p$ and a radiative decay $\gamma_{12}$. The last factor of
the first resonance, $\hbar \Omega/2$, is associated with the
emission of a photon of frequency $\omega$ into one of the vacuum
modes. Thus, the pathway of the scattering process corresponding
to the first resonance can be sketched graphically as shown in
figure \ref{Resonances-Bare-EIT-Scattering}. We also note for
later discussion at the end of this subsection that the first
resonance is of zeroth order in the coupling field, $\Omega_c$.

The second resonance has the form
\begin{eqnarray}
\rm{2^{nt} \; Resonance} &=&  \dfrac{\hbar \Omega \Omega_p}{4}
\left[\dfrac{-\Omega_c/2}{\delta_c+i(\gamma_{13}-\gamma_{12})}\right]^2
\times \nonumber\\
&
&\dfrac{1}{\delta_p+\delta_c-\delta'_c+i(\gamma_{13}+\gamma_c')}.
\label{2nd-resonance-approx-not-cleaned-Scattering}\end{eqnarray}

This contribution becomes especially large when
$\delta_p\approx-\delta_c+\delta'_c$ leading to the approximation
$\delta_c\approx -\delta_p+\delta'_c\approx -\delta_p$. In this
case, equation \ref{2nd-resonance-approx-not-cleaned-Scattering}
reduces to
\begin{eqnarray}
\rm{2^{nd} \; Resonance} &\approx& \dfrac{\hbar \Omega }{2}
\dfrac{1}{\hbar(\delta_p+i(\gamma_{12}-\gamma_{13}))} \times\nonumber\\
& & \hspace*{-0.4cm} \dfrac{\hbar \Omega_c }{2}
\dfrac{1}{\hbar(\delta_p+\delta_c-\delta'_c+i(\gamma_{13}+\gamma'_c))}
\dfrac{\hbar \Omega_c }{2} \times\nonumber\\
& &\hspace*{-0.4cm}
\dfrac{1}{\hbar(\delta_p+i(\gamma_{12}-\gamma_{13}))} \dfrac{\hbar
\Omega_p }{2}. \label{second_resonance_EIT_bare_final_version}
\end{eqnarray}

The second resonance (Fig. \ref{Resonances-Bare-EIT-Scattering})
corresponds to the absorption of the probe photon followed by
simultaneous absorption and emission of coupling field photons and
ending with the spontaneous emission of one vacuum photon.

The transition amplitude (Eq.
\ref{T_fi-Bare-EIT-before-aprox-Scattering}), after all the
approximations discussed previously, reduces to
\begin{eqnarray}
\mathcal{T}_{fi} &\approx& \dfrac{\hbar \Omega }{2}
\dfrac{1}{\hbar(\delta_p+i\gamma_{12})} \dfrac{\hbar \Omega_p
}{2}+\dfrac{\hbar \Omega }{2}
\dfrac{1}{\hbar(\delta_p+i(\gamma_{12}-\gamma_{13}))}\times \nonumber\\
& & \hspace*{-0.2cm}  \dfrac{\hbar \Omega_c }{2}
\dfrac{1}{\hbar(\delta_p+\delta_c-\delta'_c+i(\gamma_{13}+\gamma'_c)))}
\dfrac{\hbar \Omega_c }{2} \times \nonumber\\
& & \hspace*{-0.2cm}
\dfrac{1}{\hbar(\delta_p+i(\gamma_{12}-\gamma_{13}))} \dfrac{\hbar
\Omega_p }{2}.
\label{T_fi-Bare-EIT-after-aprox-Scattering}\end{eqnarray}

We stress again the fact that the requirement to have interference
is not only the existence of two or more interfering pathways but
also comparable probability amplitudes between the resonances. In
the Cascade-EIT case, the second resonance (Eq.
\ref{second_resonance_EIT_bare_final_version}) approaches the
first resonance (Eq. \ref{first_resonance_EIT_bare_final_version})
in magnitude as $\Omega_c$ becomes larger than $\gamma_{13}$, a
situation which does not violate the low saturation condition,
$\Omega^2_c /(\gamma_{12}-\gamma_{13})^2 \ll 1$.

\section{Dressed States Picture}
\label{Dressed-States-Picture-Scattering}

Earlier, we mentioned that quasistable bare states do not exist in
the Cascade-AT configuration. This makes the scattering technique
unsuitable for the analysis of the scattering process. This
difficulty can be removed in the dressed states picture, and in
particular in the low saturation limit, as we are going to show in
this section.

As our first step we describe the Cascade-AT configuration in the
dressed states picture and prove the absence of interference in
the low saturation limit. In subsection
\ref{Cascade-EIT-Scattering} we study the Cascade-EIT case and
reproduce the results found in the bare state picture (the two
pictures, of course, must be equivalent to each other).

\subsection{Cascade-AT}
\label{Cascade-AT-Scattering}

In the low saturation limit, $\Omega^2_c/\delta^2_c \ll 1 $, the
two eigenstates of the interaction Hamiltonian of the Cascade-AT
system have the form
\begin{subequations}
\begin{eqnarray}
\ket{a(N_c)} &=& \ket{1,N_c}+\dfrac{\Omega_c}{2\delta_c} \ket{2,N_c-1},\label{dressed-state-a-lsl-AT}\\
\ket{b(N_c)} &=& \ket{2,N_c-1}-\dfrac{\Omega_c}{2\delta_c}
\ket{1,N_c},\label{dressed-state-b-lsl-AT}
\end{eqnarray}\label{dressed-state-lsl-AT}
\end{subequations}
where $\delta_c = \omega_c - \omega_{12}$ and the coupling Rabi
frequency is defined as
\begin{equation}
\Omega_c \;=\; -2\mu_{12}\sqrt{\frac{\hbar\omega_c N_c
}{2\epsilon_o L^3}},
\end{equation}
and correspondingly the eigenvalues of the eigenstates (Eqs.
\ref{dressed-state-lsl-AT}) are given by
\begin{subequations}
\begin{eqnarray}
E_a &=& 0,\label{E_a-lsl-Scattering}\\
E_b &=& -\delta_c.
\end{eqnarray}
\end{subequations}

The dressed state $\ket{a(N_c)}$ (Eq.
\ref{dressed-state-a-lsl-AT}) is made of the sum of the atomic
bare state $\ket{1}$ and a correction term due to the coupling of
the atom with the field. Because the correction term is small in
the low saturation limit, the spontaneous decay rate out of level
$\ket{a(N_c)}$ is approximately equal to that of the ground state,
$\ket{1}$, i.e. the state is quasi-stable.

The same argument holds for the other dressed state,
$\ket{b(N_c)}$, which approaches to the atomic state $\ket{2}$ in
the absence of the coupling field. Based on the previous arguments
the total decay rates of the two dressed states are given by
\begin{subequations}
\begin{eqnarray}
\Gamma_a &=& 0,\\
\Gamma_b &=& W_{21}.
\end{eqnarray}
\end{subequations}

In this case the dressed state $\ket{a(N_c)}$ is quasi-stable and
can be used as the initial and final state of the scattering
process. The process begins with one photon in the probe field and
ends with one photon of energy $\hbar \omega'$ in the vacuum and
no photons in the probe field. Thus, we define the following
initial and final states
\begin{subequations}
\begin{eqnarray}
\ket{i} &=& \ket{a(N_c),(1)_p,(0)_j},\label{initial-state-AT-Scattering}\\
\ket{f} &=&
\ket{a(N_c),(0)_p,(1)_{\omega'}}.\label{final-state-AT-Scattering}
\end{eqnarray}
\end{subequations}

In this case the transition amplitude (Eq. \ref{T-fi-general-Secular}) reduces to
\begin{equation}
T_{fi}(E_i+i\eta) = \dfrac{\hbar^2\Omega'\Omega_p}{4}
\left(\dfrac{\Omega_c}{2\delta_c}\right)^2 G_{33}(E_i+i\eta),
\label{T_fi-G-AT-Scattering}\end{equation}
where we defined the Resolvent matrix element $G_{33}(E_i+i\eta)$
as
\begin{eqnarray}
G_{33}(E_i+i\eta) &=& \bra{3(N_c),(0)_p,(0)_j} \times \nonumber\\
& & G(E_i+i\eta)\ket{3(N_c),(0)_p,(0)_j},
\end{eqnarray}
and the probe and non-zero vacuum field Rabi frequencies as
\begin{subequations}
\begin{eqnarray}
\Omega' &=& -2\mu_{23}\sqrt{\frac{\hbar\omega'}{2\epsilon_o
L^3}},\\
\Omega_p &=& -2\mu_{23}\sqrt{\frac{\hbar\omega_p}{2\epsilon_o
L^3}}.
\end{eqnarray}
\end{subequations}

The intermediate state $\ket{3(N_c),(0)_p,(0)_j}$ belongs to its
own one dimensional space. In this case, the corresponding form of
equation \ref{PG(z)P-R-Secular} in a one dimensional space leads
to
\begin{equation}
G_{33}(E_i+i\eta) \;=\; \dfrac{1}{E_i+i\eta-E_3-R_{33}}.
\label{(PGP)_{33}-AT-Scattering}\end{equation}
where the matrix element $R_{33}$ of the level-shift operator is
given by
\begin{equation}
\lim_{\eta \rightarrow 0_{+}}R_{33} (E_i+i\eta) \;=\; -i\hbar
\dfrac{W_{32}+W_{31}}{2}, \label{R_33-EIT-Bare-Scattering}
\end{equation}
and $E_i$ ($E_3$) is the energy of the state $\ket{i}$
(\ket{3(N_c),(0)_p,(0)_j}), $E_i-E_3=\hbar(\delta_p+\delta_c)$,
where in the Cascade-AT case the probe detuning, $\delta_p$, is
defined as $\delta_p=\omega_p-\omega_{23}$.

After substituting equation \ref{(PGP)_{33}-AT-Scattering} into
equation \ref{T_fi-G-AT-Scattering} we obtain
\begin{equation}
T_{fi}(E_i+i\eta) \;=\;  \dfrac{\Omega_c}{2\delta_c}
\dfrac{\hbar\Omega'}{2} \dfrac{1}{\hbar\left(\delta_p + \delta_c
+i \gamma_{13}\right)}
\dfrac{\hbar\Omega_p}{2}\dfrac{\Omega_c}{2\delta_c},
\label{T_fi-final-AT-Scattering}\end{equation}
where we replaced $(W_{32}+W_{31})/2$ with the polarization decay rate $\gamma_{13}$,
which can be obtained from equation \ref{polarization_general_eq}.\\

\begin{figure}[htbp]
  \centering
\includegraphics[angle=0,width=6.0cm]{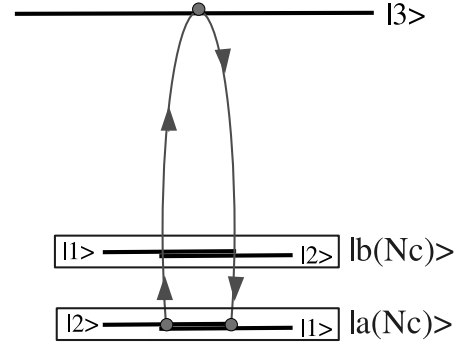}
  \caption{Resonances of the Cascade-AT configuration in the dressed states picture.}
  \label{Resonances-Dressed-AT-Scattering}
\end{figure}

The transition amplitude (Eq. \ref{T_fi-final-AT-Scattering})
consists of only one term corresponding to a single resonance.
There is only one complex number associated with the only pathway
followed by the dressed atom as it evolves from the initial state
(Eq. \ref{initial-state-AT-Scattering}) to the final state (Eq.
\ref{final-state-AT-Scattering}). This fact shows the absence of
interference in the absorption process of a probe photon in the
Cascade-AT case in the weak coupling field regime. The process
described by this transition amplitude is shown by figure
\ref{Resonances-Dressed-AT-Scattering} where we have taken the
detuning of the coupling field to be negative, a choice which
leads to a higher energy level for the state $\ket{b(N_c)}$,
relative to the state $\ket{a(N_c)}$. In addition to the
stimulated absorption and emission of photons out and into the
mode of the coupling field within the dressed state
$\ket{a(N_c)}$, the scattering process involves the absorption of
a probe photon followed by the emission of a photon in the vacuum
field via the intermediate state $\ket{3(N_c),(0)_p,(0)_j}$. It is
true that other pathways of higher order of interaction in the
coupling field exist but these processes are extremely weak and
can be ignored in the low saturation limit. We did beside the work
presented in this paper extensively check out the first higher
order process at resonance, $\delta_c =0$, and proved that it can
indeed be ignored in the low saturation limit $\Omega^2_c /
\gamma^2_{12}\ll 1$.

\subsection{Cascade-EIT}
\label{Cascade-EIT-Scattering}

The dressed states of the Cascade-EIT system are
\begin{subequations}
\begin{eqnarray}
\ket{a(N_c)} &=& \ket{2,N_c}+\dfrac{\Omega_c}{2\delta_c} \ket{3,N_c-1},\label{dressed-state-a-lsl-EIT}\\
\ket{b(N_c)} &=& \ket{3,N_c-1}-\dfrac{\Omega_c}{2\delta_c}
\ket{2,N_c}.\label{dressed-state-b-lsl-EIT}
\end{eqnarray}
\end{subequations}

We use here the same initial and final states (Eqs.
\ref{initial-state-Bare-EIT-Scattering} and
\ref{final-state-Bare-EIT-Scattering}) defined in the bare states
picture. In the low saturation limit, $\Omega^2_c/\delta^2_c \ll 1
$, and in the dressed states picture the transition amplitude (Eq.
\ref{T-fi-general-Secular}) reduces to
\begin{eqnarray}
\mathcal{T}_{fi} &=&  \dfrac{\hbar^2\Omega\Omega_p}{4} \times \nonumber\\
& & \hspace*{-0.5cm}  [\bra{a(N_c);(0)_p,(0)_j}G(E_i+i\eta)\ket{a(N_c);(0)_p,(0)_j} +\nonumber\\
& & \hspace*{-0.5cm} \dfrac{\Omega^2_c}{4\delta^2_c} \bra{b(N_c);(0)_p,(0)_j}G(E_i+i\eta)\ket{b(N_c);(0)_p,(0)_j}].\nonumber\\
\label{T-fi-Dressed-oEIT-Scattering}
\end{eqnarray}

The Resolvent matrix elements corresponding to the intermediate
states $\ket{a(N_c);(0)_p,(0)_j}$ and $\ket{b(N_c);(0)_p,(0)_j}$,
which belong to two one-dimensional subspaces, are given by
\begin{subequations}
\begin{eqnarray}
\hspace*{-0.5cm} G_{aa}(E_i+i\eta)&=&\dfrac{1}{E_i+i\eta-E_a-i\hbar\dfrac{W_{21}}{2}},\\
\hspace*{-0.5cm}
G_{bb}(E_i+i\eta)&=&\dfrac{1}{E_i+i\eta-E_b-i\hbar
\dfrac{W_{31}+W_{32}}{2}},
\end{eqnarray}
\label{G-aa-G-bb-EIT-Dressed-Scattering}\end{subequations}
where
\begin{subequations}
\begin{eqnarray}
E_a &=& 0,\\
E_b &=& -\delta_c,\\
E_i &=& \delta_p+\delta_c.
\end{eqnarray}
\end{subequations}

Thus, the transition amplitude (Eq. \ref{T-fi-Dressed-oEIT-Scattering}) takes the form
\begin{eqnarray}
T_{fi} &=& \dfrac{\hbar\Omega}{2}
\dfrac{1}{\hbar(\delta_p+i\gamma_{12})} \dfrac{\hbar\Omega_p}{2}
\;+\;\label{T_fi-EIT-Dressed-final-Scattering} \\
 & & \dfrac{\hbar\Omega}{2}
\dfrac{1}{\hbar\delta_c}\dfrac{\hbar\Omega_c}{2}
\;\dfrac{1}{\hbar(\delta_p+\delta_c+i\gamma_{13})}
\;\dfrac{\hbar\Omega_c}{2}\dfrac{1}{\hbar\delta_c}\dfrac{\hbar\Omega_p}{2}.\nonumber
\end{eqnarray}

\begin{figure}[htbp]
\centering
\includegraphics[angle=0,width=6.0cm]{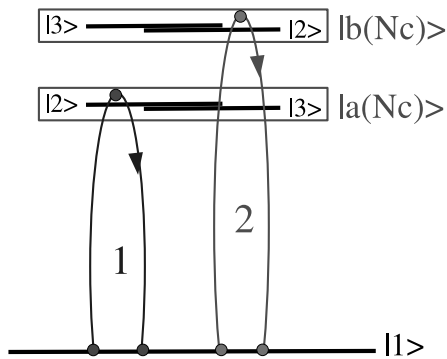}
\caption{Resonances of the Cascade-EIT configuration in the dressed states picture.}
\label{Resonances-Dressed-EIT-Scattering}
\end{figure}

The transition amplitude (Eq.
\ref{T_fi-EIT-Dressed-final-Scattering}) found here is the sum of
two terms which are associated with two resonances. Figure
\ref{Resonances-Dressed-EIT-Scattering} shows the two resonances
in the dressed states picture.

The first resonance corresponds to the excitation of the dressed
atom from level $\ket{1(N_c)}$ to level $\ket{a(N_c)}$ followed by
the decay to the ground state by spontaneous emission of one
photon into the vacuum. The second term corresponds to a process
which is very similar to the first but where the dressed atom gets
excited to the dressed state $\ket{b(N_c)}$ not $\ket{a(N_c)}$.

The transition amplitude (Eq.
\ref{T_fi-Bare-EIT-after-aprox-Scattering}), calculated in the
bare states picture in the low saturation limit, has the same
structure as the amplitude (Eq.
\ref{T_fi-EIT-Dressed-final-Scattering}) derived in the dressed
states picture under the same conditions. The physical
interpretations given for the two resonances, first resonance and
second resonance, at the end of subsection
\ref{Bare-States-Picture-lsl-Scattering} are consistent with the
previous discussion given for the transition amplitude (Eq.
\ref{T_fi-EIT-Dressed-final-Scattering}). In this case,
$\Omega^2_c/\delta^2_c \ll 1 $, the magnitude of the second
resonance in equation \ref{T_fi-EIT-Dressed-final-Scattering}
approaches the magnitude of the first resonance under the same
condition $\Omega_c > \gamma_{13}$ which was introduced in the
bare states picture.

\vspace{0.3cm}
\subsection{Conclusions}

In this section, Dressed States Picture, we clearly showed the
physical difference between the Cascade-EIT and Cascade-AT
configurations. In the weak coupling field regime, unlike the case
in the Cascade-EIT setting where the scattering process of one
probe photon can happen following either one of two distinct and
compatible pathways, in the Cascade-AT setting only one evolution
pathway dominates over the rest.

The existence of two scattering pathways of comparable
probabilities in the Cascade-EIT configuration leads to the
manifestation of interference effects. This remarkable
interference phenomenon is absent in the Cascade-AT configuration
where the atom follows only one scattering pathway in the low
saturation limit.

\section{Conclusions}
\label{Conclusions}

By applying the scattering technique we studied the Cascade-EIT
configuration in the bare and dressed states pictures. The
Cascade-AT configuration was studied only in the dressed states
picture for reasons that were stated in section
\ref{Bare-States-Picture-Scattering}. In the Cascade-EIT case we
showed that the transition amplitude is the sum of two complex
numbers associated with two resonances. These resonances
correspond to interfering scattering pathways which we described
in the bare states picture (Fig.
\ref{Resonances-Bare-EIT-Scattering}), and in the dressed states
picture (Fig. \ref{Resonances-Dressed-EIT-Scattering}) in the weak
field regime. The transition amplitude corresponding to the
Cascade-AT case was derived in the low saturation limit, and in
the dressed states picture. The calculated amplitude was
associated with one resonance presented in figure
\ref{Resonances-Dressed-AT-Scattering}. The domination of only one
scattering pathway eliminates the possibility of interference
effects in the Cascade-AT configuration.

We also showed that after considering the low saturation limit,
the derived transition amplitude (Eq.
\ref{T_fi-Bare-EIT-after-aprox-Scattering}) for the Cascade-EIT
configuration found in the bare states picture (Sec.
\ref{Dressed-States-Picture-Scattering}) approaches analytically
the amplitude (Eq. \ref{T_fi-EIT-Dressed-final-Scattering}) found
in the Dressed states picture in the low saturation limit as well.

By studying the two cascade configurations, Cascade-EIT and
Cascade-AT, we revealed two different coupling field regimes. We
explored the weak field regime in detail and proved that unlike
the AT effect EIT persists in the low saturation limit due to the
existence of interference between competing pathways.


\section{Acknowledgments}

This work was conducted under the guidance and supervision of the
late professor Lorenzo Narducci whom I dearly admire and miss. I
would also like to thank Dr. Frank Narducci for critical reading
of the manuscript and his genuine feedback.


\end{document}